# Quantum Proof of Work with Parametrized Quantum Circuits

**Mikhail Y. Shalaginov[1] and Michael Dubrovsky[2]**

[1]Department of Materials Science & Engineering, Massachusetts Institute of Technology, Cambridge, MA, USA, mys@mit.edu
[2]PoWx, Cambridge, MA, USA, mike@powx.org

**ABSTRACT**

Despite all the progress in quantum technologies over the last decade, there is still a dearth of practical applications for quantum computers with a small number of noisy qubits. The effort to show quantum supremacy has been largely focused on demonstrating computations that cannot be accomplished on a classical computer at all, a difficult and controversial target. Quantum advantage (a speedup over classical computers) is a more practical milestone for today's modest quantum processors. In this work, we proposed a scheme for quantum-computer compatible proof of work (cryptographic mechanism used in Bitcoin mining) and verified it on a 4-qubit superconducting quantum node.

**INTRODUCTION**

Public cryptocurrency networks, such as Bitcoin, maintain an immutable, decentralized ledger of transactions in electronic currency. Albeit the past attempts to establish e-cash systems, the Bitcoin's architecture (outlined in the original Nakamoto whitepaper [1]) was the first one to solve the double-spend [2] and the Sybil attack [3] problems through the clever use of Hashcash [4] and Proof of Work (PoW) [5]. The continued reliability and relative stability of the Bitcoin network throughout its massive growth over the past decade has made Bitcoin a real contender for the status of "world currency."

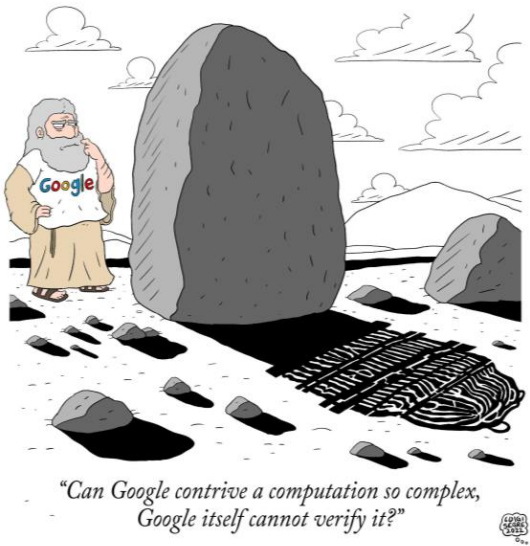

**Figure 1**. Cartoon illustrating the quantum supremacy version of the omnipotence paradox.

Bitcoin's security hinges on PoW, the main computational task in cryptocurrency mining. A new PoW protocol leveraging the power of quantum computers (QCs) may redirect some portion of the billions of dollars currently spent yearly on cryptocurrency mining with traditional ASICs and GPUs to quantum computing R&D. This trend has an additional benefit of decreasing the electricity consumption caused by conventional Bitcoin mining which already surpasses that of several European countries [6]. Here we demonstrate a quantum PoW (qPoW) protocol that can be accelerated with available quantum computers while maintaining all the required security properties. qPoW borrows some of its architecture from the earlier proposed optical PoW (oPoW) which relies on the matrix-multiplication speed-up of analog optical computers [7].

A famous paradox asks: "Can God make a stone so heavy that he himself cannot lift it?"[8] In 2019 Google team announced that they built a quantum processor so complex that Google itself could not directly verify the computed results [9]. The team claimed quantum supremacy by relying on the sample statistics collected from the pseudo-random 53-qubit circuit to prove that indeed produced quantum computations could not be verified within a reasonable time (> 10,000 years) by the most powerful classical computer. This assertion was later challenged by IBM and others [10], [11].

Approaching the opportunities of quantum computers from a different side, we wondered whether there is a quantum computation task that is already possible to run on existing quantum computers, difficult to compute classically (though not impossible, and therefore verifiable), and sufficiently deterministic to be used in PoW. Such quantum task can be embedded in a PoW blockchain which is advantageous to mine with quantum computers and yet possible to verify with classical computers.

## RESULTS

**Practical considerations for qPoW.** To utilize Hashcash mining, qPoW must contain the same cryptographic security as SHA256, the secure hash algorithm used in Bitcoin. This requirement is achieved by using a similar cryptographic architecture as in optical PoW [7] with a modification to the algorithm that replaced an optical computation block with a novel quantum computation block (see Fig. 2A). When a miner finds a satisfactory output, it must be possible for others to verify the computation using a classical computer, i.e., the qPoW result obtained with a quantum computer must agree with a simulator to be considered "valid." Hence, if someone runs qPoW on a quantum computer, its results must agree with classical simulations often enough to be useful. The accuracy may be decreased to the level when a quantum processor still has a noticeable computational speedup in comparison to a classical computer. In general, there is no penalty for producing invalid solutions besides time wasted running the quantum node. Note that the qPoW protocol has to be performed many times to mine a block by varying successive data inputs. To verify a mined block, one only needs to run a simulation cycle once. Since every node in the network has to go through block verification, this procedure should be relatively inexpensive. Finally, to make it economically practical to perform qPoW on quantum nodes, the quantum systems must give a sufficient speedup (×100 – 10,000) of the algorithm relative to classical ones.

**Quantum circuit ansatz.** The 256-bit-string produced by the first SHA3 function (the first orange block in Fig. 2A) is split into 4-bit substrings. Afterwards, each substring number is turned into a decimal, which is further encoded to a rotational angle of the quantum-circuit gates $R_x$, $R_z$, and $cR_x$ (Fig. 2B). For example, a 4-bit substring '0101' (9 in a decimal format) will be converted into a rotation angle of $9·\pi/8$ of the specific rotational gate. The ansatz architecture here was adopted from [12] as an ansatz of a highly expressible quantum circuit, which is known to produce non-uniform output state distributions.

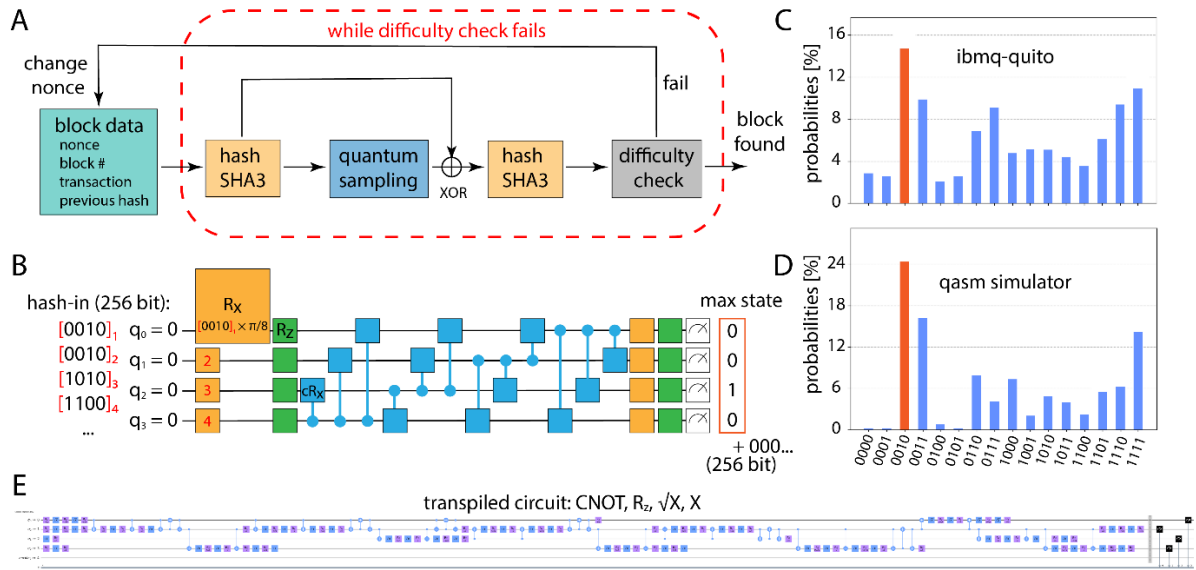

**Figure 2. Quantum Proof-of-Work. A**: Schematic of a HeavyHash PoW with a quantum sampling module. **B:** Detailed illustration of a quantum sampling algorithm that encodes a pre-hashed 256-bit string (broken into quads) as angles of the rotational gates $R_x$, $R_z$, $cR_x$ and outputs a 4-bit string of the most probable state. **C, D**: Exemplary output histograms of the measured quantum states after 20,000 shots produced by **(C)** an IBM quantum processor 'ibmq_quito' and **(D)** by qasm simulator. **(E)** Transpiled version of an exemplary quantum ansatz.

**Example of a qPoW computation.** The input data ('text') is a concatenated string containing: nonce, transaction information and a hash string from a previous block. A nonce is a 32-bit randomly generated integer; transaction information is a ledger entry, e.g. 'Schroedinger paid Einstein 1 qBTC'. By SHA3 algorithm the 'text' (e.g. '4Schroedinger paid Einstein 1 qBTC04ca1a782621a440d03b5d87ecff8b68e2cc6124f57957b49a76bca91dede3a81') is transformed into a 256-bit hash-string ('e1e5575da3a9e86da135552facddcc1ff44dd26502d0bc2b22961383f8b187ca') that is then pushed through the qPoW stages (Fig. 2A, B). After the second hash module, we have a 256-bit hash-string as an output ('f307b3db12a649563831e3e1328c3c7a5b15ee541afaab563727cb992cf9d1ca'). If the output satisfies the conditions set by the task difficulty, i.e. the output string starts from a certain number of zeros, then the block is reported to be successfully mined. Otherwise, the output string is discarded and a new nonce is randomly generated for the next mining cycle. When the block is finally found (i.e., the difficulty check is passed), it goes through a verification test performed by a quantum simulator ran on a classical computer.

**qPoW on IBM quantum node vs Qiskit Aer simulator**. We tested a 4-qubit ansatz on a publicly available 5-qubit processor (ibmq_quito) and found that the runtime of qPoW on ibmq_quito is nearly 100 times longer than on the simulator (qasm). Sometimes ibmq_quito and qasm disagree and output distinct maximum quantum states. We assume this disagreement originates from the computational errors on the quantum node, primarily induced by CNOT gates and readout operations. To run a quantum circuit on real quantum hardware, the circuit needs to go through transpilation which results in a circuit adapted to the quantum hardware topology and composed of native gates (e.g., ID, RZ, SX, NOT, CNOT). The transpiled 4-qubit quantum circuit (see Fig. 2E) contains on average 40 CNOTs. According to the information provided by IBM, the average error rate of a CNOT gate on ibmq_quito is 1%, therefore, the overall circuit accuracy is expected to be 67%. In a statistical set of 100 iterations on ibm_quito and qasm, we experimentally recorded a coincidence probability of 69%. Values of the measured accuracy and the number of CNOTs in a transpiled circuit for other numbers of qubits are provided in Table 1.

**Table 1.** Statistics of the pseudo-random quantum-circuit performance used in qPoW.

| # of qubits | 2 | 3 | 4 | 5 |
|---|---|---|---|---|
| measured accuracy, % | 94 | 71 | 69 | 35 |
| # of CNOTs (transpiled circuit) | 3.7 | 17.5 | 40.4 | 90.7 |
| simulator vs QC time ratio, 1e-3 | 6 | 8 | 8.4 | 10.5 |

**Demonstration of a primitive blockchain based on qPoW.** We showcased the construction of a block-chain composed of 5 consecutive blocks by running qPoW on 4 qubits of ibmq_quito with the difficulty of finding a hex string starting with a single '0' character. A chain of 5 blocks was successfully constructed. During the procedure two blocks out of 7 that were found by the quantum computer could not pass verification performed on a qasm simulator. Hence, the success rate is 71%, which is consistent with the prior measured accuracy of the 4-qubit quantum computer.

## DISCUSSION
**Requirements for a quantum computer to outperform a simulator in qPoW task.** On a publicly available 4-qubit quantum computer we have successfully demonstrated qPoW based mining, however, the computational time was almost 100 times longer than on a qasm simulator (Table 1). This observation raises a question: how many qubits are needed to show that a noisy quantum computer has a noticeable boost in qPoW mining, e.g., ×10 – 10000 than the existing classical computers. To find the answer, we first ran the qPoW task with the generalized n-qubit quantum ansatz (Fig. 2B) on a qasm simulator (running on a regular desktop computer) to study the dependence of runtime $t$ vs number of qubits $n$. As shown in Fig. 3A, starting from 15 qubits the task complexity for a classical computer (blue dots) grows exponentially with $n$ as $t = 10^{0.33n}$ (blue dashed line in Fig. 3A). The complexity for a quantum processor is assumed to be proportional to the quantum circuit depth (red dashed line in Fig. 3A), i.e., polynomial dependence ~ $n^2$ [12]. Hence, for $n > 20$ qubits (area highlighted in gray) the quantum computer is predicted to outpace the qasm simulator.

This is not the end of the story. Quantum computers are noisy, i.e., each of the two-qubit gates induces an error on the order of 1%, measuring each qubit also leads to an error of nearly 2%. Here comes the tradeoff: by increasing $n$, the number of qubits in the ansatz (increasing qPoW complexity), the accuracy of QC sampling drops. Therefore, we define quantum advantage as a product of speed ratio (inverse ratio of qasm and QC runtimes) and the QC accuracy. The speed ratio is evaluated from the extrapolated blue and red curves in Fig. 3A. The accuracy for the $n$-qubit sampling ansatz has been estimated as $(1 - E_{\text{CNOT}})^{\text{CNOTs}} \cdot (1 - E_{\text{readout}})^n$, where $E_{\text{CNOT}}$ and $E_{\text{readout}}$ are the realistic errors of one CNOT gate (1%) and one readout operation (2%); CNOTs – number of CNOTs in the circuit. Here we pick CNOT's parameters since they are the native two-qubit gates of ibm_quito, and in the end all the other multi-qubit gates are eventually translated into them. To improve the accuracy, it is highly recommended to use primarily the set of the native gates (ID, RZ, SX, NOT, CNOT), otherwise transpilation procedure will significantly increase the number of CNOTs, at least by an order of magnitude (for illustration see Fig. 2B, E), and plummet the accuracy. In this case, we consider the best-case scenario of a hypothetical non-transpiled quantum circuit, the number of two-qubit gates ~ $n^2 - n$; the resulting accuracy curve is depicted with a black dashed line and the resulting quantum advantage (product of speed ratio and accuracy) with a red line in Fig. 3B. According to our estimates, the existing superconducting quantum computer should have at least ~30 qubits to perform qPoW with a "quantum advantage" over the qasm simulator for the specific parametrized quantum sampling algorithm. Accuracy, and therefore advantage can be further improved with a shallower quantum ansatz and careful consideration of the specific QC quantum connectivity and topology.

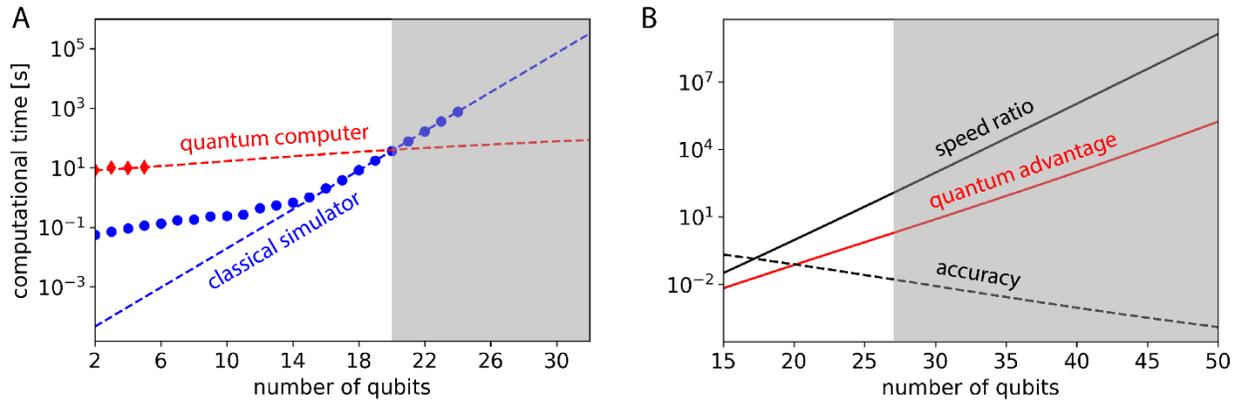

**Figure 3. Minimum viable quantum advantage. A:** Dependences of computational time $t$ of the generalized quantum ansatz (Fig. 2B) on the number of qubits $n$. The task complexity increases with the number of qubits, from 20 qubits and above the quantum computer outspeeds the classical one. Dashed blue line – exponential fit for $n \geq 15$: $t = 10^{0.33n - 5}$, dashed red line – polynomial curve: $t = 0.07(n^2+3n) + 7.5$. **B:** Projected trends of the speed ratio (inverse time ratio of the dashed curves in **A**); quantum computer accuracy, which is affected by the number of two-qubit gates with each having an error rate of 1% and qubit readout error of 2%; and quantum advantage (product of the speed ratio and the accuracy).

In conclusion, we devised a hash-based proof-of-work mining algorithm with an embedded quantum task (qPoW). The quantum task was to find a maximum state of a pseudo-random quantum circuit with parametrized rotational gates. By using a 4-qubit quantum computer we demonstrated the possibility of constructing a qPoW-based primitive blockchain. Our estimates predict that a noisy superconducting quantum computer with more than 30 qubits would reach the level of a minimal viable quantum advantage and start outperforming in mining a classical computer with a quantum simulator. qPoW can also leverage other types of computationally hard problems, such as approximation of many-body Hamiltonian states [13], boson sampling [14], combinatorics on graphs [15], etc.

**Code availability:** https://github.com/shalm/quantum-Proof-of-Work

**Acknowledgements**
The authors would like to thank Bogdan Penkovsky, Brian Dellabetta, Igor Balla, and Marshall Ball for helpful discussions as well as the organizers of the MIT Quantum Hackathon iQuHACK 2022, the event where the first qPoW code version was developed and demonstrated.